\documentclass{article}

\PassOptionsToPackage{numbers, compress}{natbib}
\usepackage[final]{neurips_2024_ml4ps}

\usepackage[utf8]{inputenc} % allow utf-8 input
\usepackage[T1]{fontenc}    % use 8-bit T1 fonts
\usepackage{hyperref}       % hyperlinks
\usepackage{url}            % simple URL typesetting
\usepackage{booktabs}       % professional-quality tables
\usepackage{amsfonts}       % blackboard math symbols
\usepackage{nicefrac}       % compact symbols for 1/2, etc.
\usepackage{microtype}      % microtypography
\usepackage{xcolor}         % colors

\usepackage{amsmath}
\usepackage{amssymb}
\usepackage{siunitx}
\usepackage{bm}
\usepackage{graphicx}
\usepackage{cleveref}
\usepackage[normalem]{ulem}

\crefname{equation}{Eq.}{Eqs.}
\crefname{figure}{Fig.}{Figs.}
\crefname{table}{Tab.}{Tabs.}

\title{Differentiable Voxel-based X-ray Rendering\\Improves Sparse-View 3D CBCT Reconstruction}

\author{
Mohammadhossein Momeni$^{1,2}$\thanks{Equal contribution. Implementation available at \url{https://github.com/hossein-momeni/DiffVox}.} \quad Vivek Gopalakrishnan$^{1,2}$\footnotemark[1] \\ \textbf{Neel Dey}$^2$ \quad \textbf{Polina Golland}$^2$ \quad \textbf{Sarah Frisken}$^{1}$ \\
$^1$Brigham and Women's Hospital \quad $^2$MIT CSAIL
}

\begin{document}

\maketitle

\begin{abstract}
We present DiffVox, a self-supervised framework for Cone-Beam Computed Tomography (CBCT) reconstruction by directly optimizing a voxelgrid representation using physics-based differentiable X-ray rendering. Further, we investigate how the different implementations of the X-ray image formation model in the renderer affect the quality of 3D reconstruction and novel view synthesis. When combined with our regularized voxel-based learning framework, we find that using an exact implementation of the discrete Beer-Lambert law for X-ray attenuation in the renderer outperforms both widely used iterative CBCT reconstruction algorithms and modern neural field approaches, particularly when given only a few input views. As a result, we reconstruct high-fidelity 3D CBCT volumes from fewer X-rays, potentially reducing ionizing radiation exposure and improving diagnostic utility.
\end{abstract}

\section{Introduction}
CBCT reconstruction estimates internal 3D structure from multiple 2D X-ray projection images, typically acquired in a circular orbit about the subject~(\cref{fig:renderer}A). Current analytical and iterative solvers for CBCT reconstruction~\cite{feldkamp1984practical, landman2023krylov, gregor2008computational} fail when reconstructing volumes from a limited number of 2D X-rays, typically prescribed to minimize radiation exposure to patients. To this end, several neural field-based methods aim to fill the need for sparse-view CBCT reconstruction~\cite{zha2022naf, ruckert2022neat, lin2023learning}. However, these methods rely on computationally intensive Transformer- or MLP-hashgrid hybrid architectures~\cite{muller2022instant}, which introduce numerous additional hyperparameters and substantially increase reconstruction times. Furthermore, previous neural fields for CBCT reconstruction have predominantly been evaluated on synthetic X-ray datasets~\cite{cai2024structure}, making it unclear if these architectures scale to real X-ray images. Despite their high performance on synthetic datasets, neural field-based CBCT reconstruction methods have not seen widespread adoption in real-world tasks.

Instead of parameterizing the unknown volume as a neural network as in previous work~\cite{zha2022naf, ruckert2022neat, lin2023learning}, we propose to perform sparse-view reconstruction by directly optimizing a regularized voxelgrid representation~\cite{sun2022direct} of a CBCT scan. DiffVox, our self-supervised reconstruction framework, has few tunable hyperparameters and is driven by a physics-based differentiable X-ray renderer~\cite{gopalakrishnan2022fast}, enabling the fusion of popular voxel-based regularizers (e.g., total variation norm~\cite{rudin1992nonlinear}) with gradient-based optimizers (e.g., Adam~\cite{kingma2014adam}). We further investigate how different X-ray image formation models affect the quality of CBCT reconstruction: Siddon's method~\cite{siddon1985fast}, an exact model of the first-order effects in X-ray physics, and trilinear interpolation, a fast approximation of these phenomena. Whereas most existing reconstruction techniques use trilinear interpolation, we find that optimization with Siddon's method leads to the highest quality reconstructions and renderings in the sparse-view regime. We evaluate our methods, and many traditional and neural field baselines, on an open dataset of more than 150,000 raw X-ray images from 42 subjects and find that differentiable voxelgrid optimization with DiffVox outperforms existing CBCT reconstruction algorithms. 

\section{Methods}

\begin{figure}[t]
    \includegraphics[width=\textwidth]{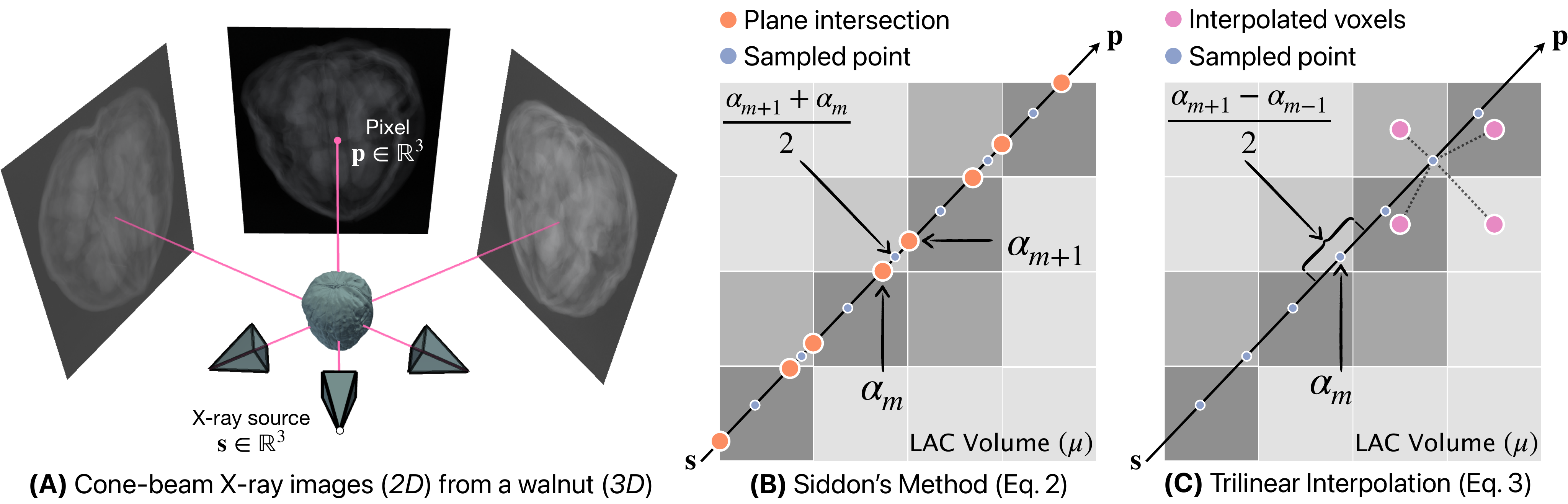}
    \caption{(\textbf{A}) In sparse-view CBCT reconstruction, a small number of X-ray images are acquired in a circular orbit about a subject. We compare two implementations of the X-ray image formation model for reconstruction via differentiable rendering: (\textbf{B}) Siddon's method and (\textbf{C}) trilinear interpolation.}
    \label{fig:renderer}
\end{figure}

\textbf{Image Formation Model (\cref{fig:renderer}).} 
\label{sec:image-formation-model}
Let $\vec{\mathbf r}(\alpha) = \mathbf s + \alpha (\mathbf p - \mathbf s)$ with $\alpha \in [0, 1]$ be a beam of photons cast through a heterogeneous medium from an X-ray source $\mathbf s \in \mathbb R^3$ to a pixel on the detector $\mathbf p \in \mathbb R^3$. An X-ray image $I$ quantifies the energy-weighted fraction of attenuated incident photons for every pixel in the detector, with the negative log intensity at $\mathbf p$ being governed by the Beer-Lambert law:
\begin{equation}
    \label{eq:beer-lambert}
    I_{\bm \mu}(\vec{\mathbf r}) 
    = \int_{\mathbf x \in \vec{\mathbf r}} \bm \mu (\mathbf x) \mathrm d \mathbf x
    = \int_0^1 \bm \mu\big( \vec{\mathbf r}(\alpha) \big) \| \vec{\mathbf r}'(\alpha) \| \mathrm d\alpha
    = \| \mathbf p - \mathbf s \| \int_0^1 \bm \mu \big(\mathbf s + \alpha (\mathbf p - \mathbf s) \big) \mathrm d\alpha \,,
\end{equation}
where $\bm \mu : \mathbb R^3 \to [0, \infty)$ represents the medium's linear attenuation coefficients (LACs), a physical property proportional to the density at every point in space. 
In all practical settings, $\bm \mu$ is discretized onto a finite-resolution voxelgrid. Therefore, CBCT reconstruction via differentiable rendering requires numerical methods to compute \cref{eq:beer-lambert} over a discrete volume of optimized LACs, denoted as $\hat{\bm \mu}$, in a manner that is differentiable with respect to $\hat{\bm\mu}$. 

\textbf{Differentiable X-ray Rendering.}
The first integration technique we consider is Siddon's method~\cite{siddon1985fast}, which exactly computes a discretized version of \cref{eq:beer-lambert} as the sum of the LAC in every voxel on the path of $\vec{\mathbf r}$, weighted by the intersection length of  $\vec{\mathbf r}$ with each voxel:
\begin{equation}
    \label{eq:siddon}
    I_{\hat{\bm \mu}}(\vec{\mathbf r}) = \| \mathbf p - \mathbf s \| \sum_{m=1}^{M-1} \hat{\bm \mu} \left [ \mathbf s + \frac{\alpha_{m+1} + \alpha_m}{2} (\mathbf p - \mathbf s) \right] (\alpha_{m+1} - \alpha_m) \,,
\end{equation}
where $\{\alpha_1, \dots, \alpha_M\}$ parameterize the intersection of $\vec{\mathbf r}$ with the parallel planes comprising $\hat{\bm\mu}$, and $\hat{\bm \mu}[\cdot]$ is an indexing operation that returns of the LAC of the intersected voxel (\cref{fig:renderer}B). It was previously shown that \cref{eq:siddon} can be implemented in a completely differentiable manner~\citep{gopalakrishnan2022fast}. Instead of computing every plane intersection, which scales cubically with the resolution of $\hat{\bm\mu}$, we propose to approximate the Beer-Lambert law using interpolatory quadrature via the rectangular rule:
\begin{equation}
    \label{eq:trilinear}
    I_{\hat{\bm \mu}}(\vec{\mathbf r}) \approx \| \mathbf p - \mathbf s \| \sum_{m=1}^{M-1} \hat{\bm \mu} \left [ \mathbf s + \alpha_m (\mathbf p - \mathbf s) \right] \frac{(\alpha_{m+1} - \alpha_{m-1})}{2} \,,
\end{equation}
where $\{\alpha_1, \dots, \alpha_M\}$ parameterize $M$ evenly spaced points along $\vec{\mathbf r}$ and $\hat{\bm \mu}[\cdot]$ represents trilinear interpolation (\cref{fig:renderer}C). Trilinear interpolation is linear in $M$ and thus faster than Siddon's method.

\textbf{Differentiable CBCT Reconstruction.}
Let $\mathcal R$ represent the set of all photon beams cast in a set of 2D X-ray images of a subject, fully specifying the acquisition geometry. For each beam $\vec{\mathbf r} \in \mathcal R$, let $I_{\bm \mu}(\vec{\mathbf r}) \in [0, \infty)$ correspond to its ground truth pixel intensity. To reconstruct the unknown volume, we directly optimize the voxelgrid $\hat{\bm \mu}$ with the following photometric loss function:
\begin{equation}
    \label{eq:loss}
    \mathcal L(\hat{\bm \mu}) = \frac{1}{|\mathcal R|}\sum_{\vec{\mathbf r} \in \mathcal R} \| I_{\bm \mu}(\vec{\mathbf r}) - I_{\hat{\bm \mu}}(\vec{\mathbf r}) \|_1 + \lambda_{\mathrm{TV}} \mathrm{TV}(\hat{\bm \mu}) \,,
\end{equation}
where $\|\cdot\|_1$ is the L1 norm and $\mathrm{TV}(\cdot)$ is the total variation norm, a commonly used regularizer that encourages reconstructed volumes to be piecewise constant~\cite{rudin1992nonlinear}. Note that we implement $\mathbf I_{\hat{\bm\mu}}(\cdot)$ as a differentiable renderer specifically to minimize $\mathcal L$ using gradient-based optimization.

\section{Experiments}
\label{sec:experiments}

\begin{figure}[t]
    \includegraphics[width=\textwidth]{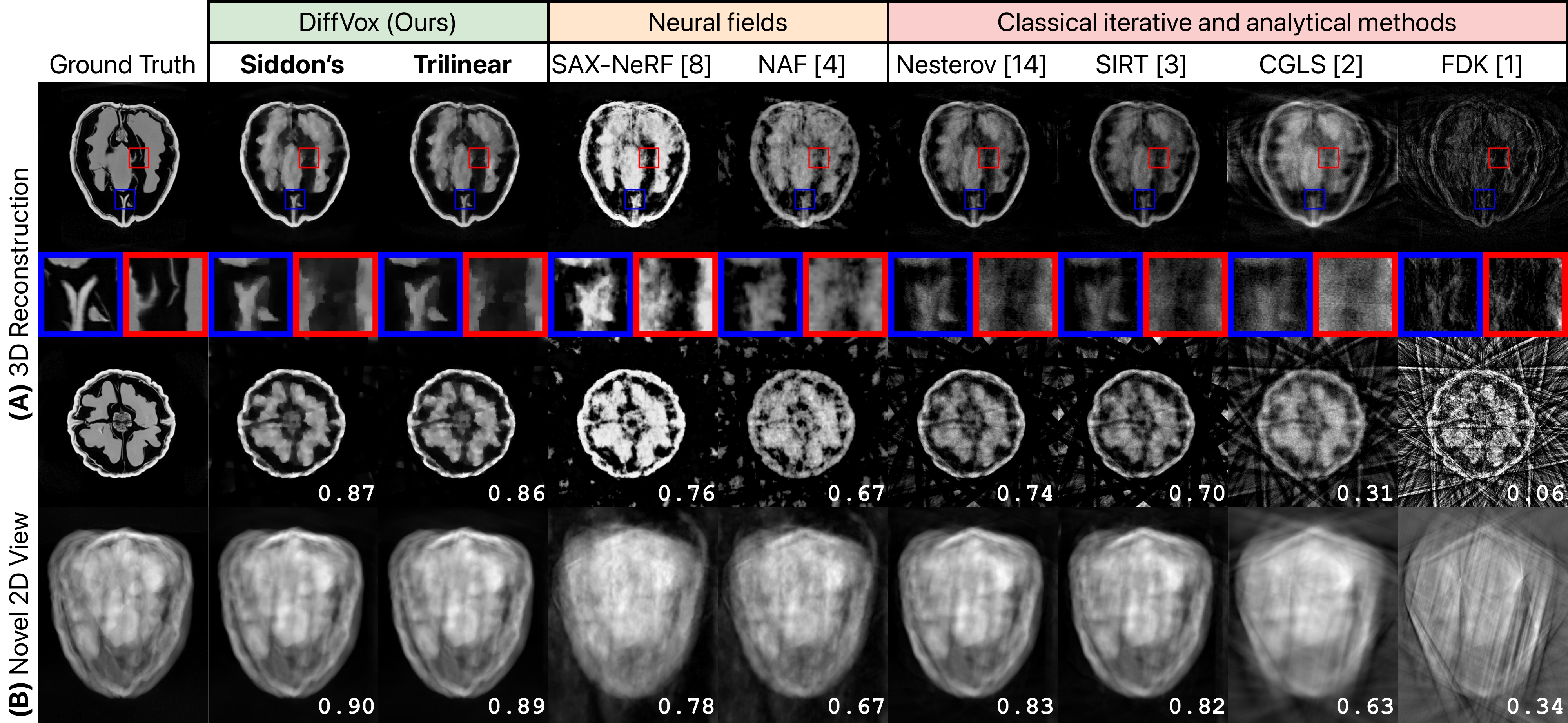}
    \caption{(\textbf{A}) CBCT reconstructions of an exemplar walnut from the test set using 15 input views. The Structural Similarity Index Measure (SSIM) for each 3D reconstruction is annotated. Blue insets highlight where our methods outperform the baselines, with sharper boundaries and fewer artifacts. Red insets indicate areas where all methods struggle, particularly in reconstructing thin structures. (\textbf{B}) Novel views rendered from these estimated volumes are compared to a ground truth X-ray image not seen during reconstruction and similarly annotated with SSIM. Novel views were rendered using the forward model implemented in each method.}
    \label{fig:qualitative}
\end{figure}

\textbf{Dataset.}
Commercial CT scanners typically do not provide access to raw projection data. Therefore, most CBCT reconstruction methods are evaluated on artificial X-ray images simulated from existing 3D scans using computational models~\cite{zha2022naf, cai2024structure}. Instead, we evaluated our proposed methods using an open dataset of real high-resolution X-ray images acquired from 42 walnuts~\cite{der2019cone}. For each walnut, 3,600 X-ray images ($972 \times 768$ pixels with \SI{0.1496}{\mm} isotropic spacing) were acquired across three circular orbits at different heights. Ground truth CBCT volumes were reconstructed using all images to produce high-resolution $501 \times 501 \times 501$ volumes at \SI{0.1}{\mm} isotropic spacing. At test-time, we reconstructed CBCT volumes of equal resolution using a sparse subset of X-ray images from only the middle orbit to replicate a typical cone-beam acquisition~(\cref{fig:qualitative}A). In addition to comparing our 3D reconstructions to the high-resolution ground truth, we also measure how similar novel 2D views rendered from our reconstructed volumes were to real 2D X-ray images from the high and low orbits, which are not seen during reconstruction~(\cref{fig:qualitative}B). All hyperparameters were tuned with a grid search over two subjects, and the remaining 40 were used as a held-out test set.

\textbf{Implementation Details.} 
We initialize $\hat{\bm\mu}$ as a voxelgrid of all zeros. As LACs are physically constrained to be nonnegative, we apply the Softplus activation function ($\beta=20$) to $\hat{\bm\mu}$ prior to rendering. To optimize $\hat{\bm\mu}$, we use the Adam optimizer with an initial learning rate of $1$ that linearly decays to $0$ over 50 iterations. In each batch, rays from all projection images are randomly sampled without replacement with uniform probability. The batch size for each renderer was set to the maximum allowable under memory constraints, i.e., $|\mathcal R| = 5.5 \times 10^5$ for Siddon's method and $|\mathcal R| = 1.8 \times 10^6$ for trilinear interpolation. Finally, we set $\lambda_{\mathrm{TV}} = 25$ for Siddon's method and $\lambda_{\mathrm{TV}} = 15$ for trilinear interpolation. $M=500$ points per ray were sampled for trilinear interpolation.

\textbf{Baselines.}
We compare against GPU-accelerated implementations of four classical CBCT reconstruction algorithms from the ASTRA Toolbox~\cite{van2016fast}: Feldkamp-Davis-Kress (FDK)~\cite{feldkamp1984practical}, Conjugate Gradient Least Squares (CGLS)~\cite{landman2023krylov}, Simultaneous Iterative Reconstruction Technique (SIRT)~\cite{gregor2008computational}, and Nesterov-Accelerated Gradient Descent (Nesterov)~\cite{der2019cone}. Based on grid searches to maximize their performance, we ran CGLS for 20 iterations, SIRT for 500 iterations, and Nesterov for 50 iterations. Additionally, we evaluated two contemporary neural field approaches for CBCT reconstruction: Neural Attenuation Fields (NAF)~\cite{zha2022naf}, an MLP-hashgrid hybrid, and Structure-Aware X-ray Neural Radiodensity Fields (SAX-NeRF)~\cite{cai2024structure}, a Transformer-hashgrid hybrid, which recently reported state-of-the-art accuracy in sparse-view CBCT reconstruction.

\textbf{Results.}
We tested the performance of DiffVox and relevant baselines in the sparse-view regime, reconstructing 40 walnuts with the number of input views ranging from $\{5, 10, 15, \dots, 60\}$. To evaluate the quality of these 3D reconstructions and subsequent novel 2D views, we computed the following image fidelity scores: Structural Similarity Index Measure (SSIM), Peak Signal-to-Noise Ratio (PSNR), Pearson's Correlation Coefficient (PCC), and Mean Squared Error (MSE). 
Additionally, runtimes for five subjects were measured using a single NVIDIA RTX A6000 GPU.

\begin{figure}[t!]
    \includegraphics[width=\textwidth]{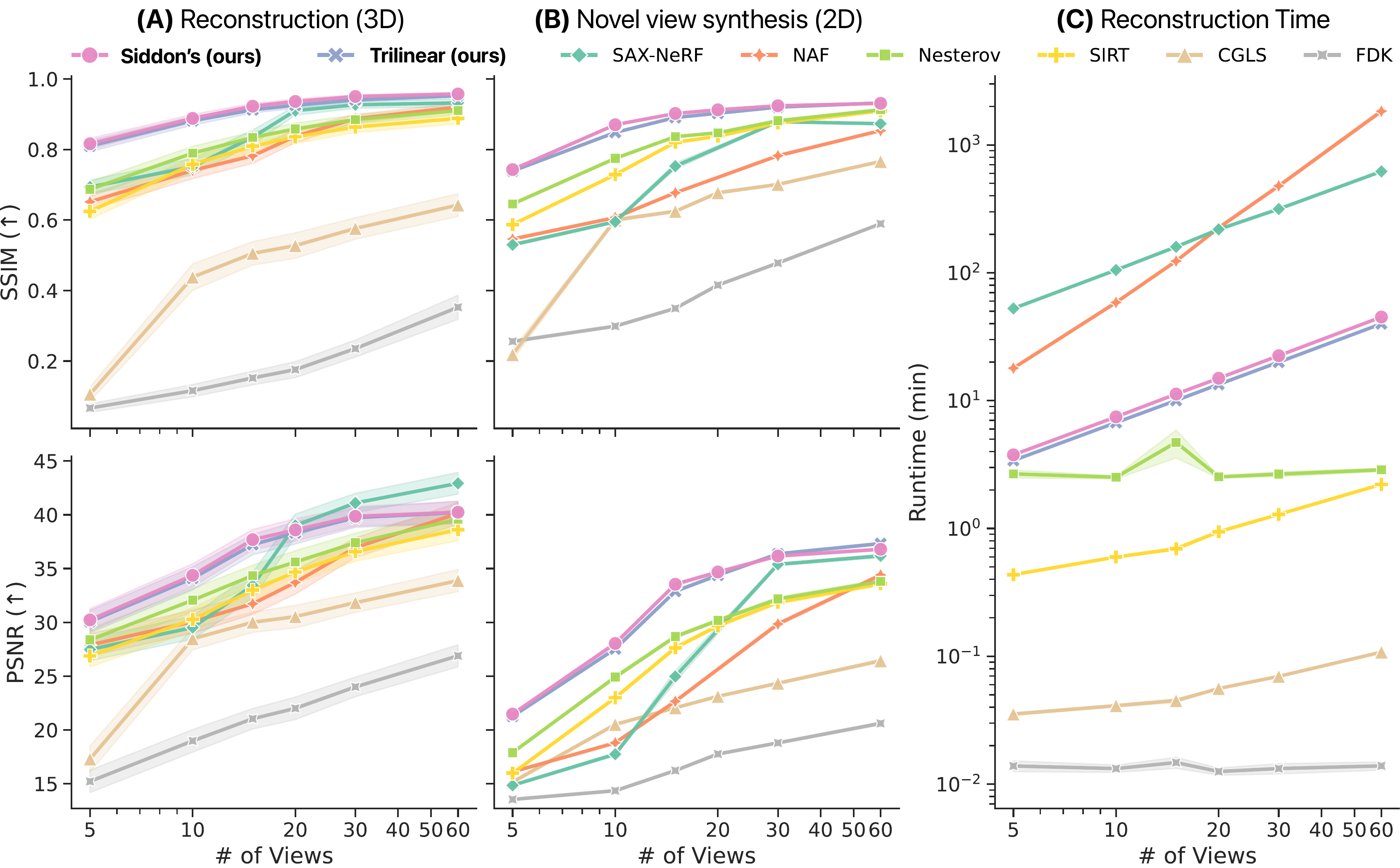}
    \caption{Quality of (\textbf{A}) reconstructed 3D volumes and (\textbf{B}) novel 2D views rendered from these volumes, and (\textbf{C}) reconstruction runtimes over a range of input views. DiffVox, using either Siddon's method or trilinear interpolation as the forward model, achieves the highest quality reconstructions and renderings, with particularly appreciable gains in the sparsest-view settings. Error bars are plotted using the standard error (\textit{se}).}
    \label{fig:quantitative}
    \vspace{-1em}
\end{figure}

In sparse-view settings, our voxel-based differentiable rendering approach outperforms classical and neural CBCT reconstruction techniques across all image quality metrics~(\cref{tab:quant}). Visualized slices of reconstructions from 15 input views in \cref{fig:qualitative}A show that our methods successfully recover the highest quality 3D volumes. In comparison, previous methods suffer from stereotypical sparse-view artifacts, such as floaters and aliasing artifacts. Consequently, the novel views rendered from our reconstructed volumes have much higher fidelity~(\cref{fig:qualitative}B). When performing reconstructions from 30 and 60 views, SAX-NeRF achieves a higher PSNR than our methods~(\cref{fig:quantitative}A, \textit{bottom}). However, across all other number of views and performance metrics, DiffVox outperforms all baseline CBCT reconstruction methods in both reconstruction and novel view synthesis tasks~(\cref{fig:quantitative}A,B).

In both 3D reconstruction and 2D novel view synthesis, Siddon's method provides a modest improvement over trilinear interpolation~(\cref{fig:quantitative}A,B) while incurring a slightly higher runtime due to its increased complexity~(\cref{fig:quantitative}C). In particular, the advantage of Siddon's method is evident when rendering 2D projections in the sparse-view setting (5-20 views in~\cref{fig:quantitative}B). In this regime, novel views rendered from volumes reconstructed with Siddon's method have consistently higher fidelity than those produced by either trilinear interpolation or any other baseline method.

For both of our differentiable rendering methods, regularization via TV norm leads to high-quality reconstructions. However, as the ablation study in \cref{tab:quant} shows, removing the TV norm has a far greater impact on Siddon's method than on trilinear interpolation. This is because, while Siddon's method exactly computes the line integral in the Beer-Lambert law (which leads to higher fidelity novel views), trilinear interpolation allows rendered pixel intensities to be influenced by voxels not directly intersected by $\vec{\mathbf r}$ (\cref{fig:renderer}C). This is advantageous as it enables ground truth 2D pixels to provide self-supervision to many more 3D voxels in the reconstruction, preventing the overfitting observed in Siddon's method. Furthermore, this behavior is an approximation of second-order effects such as X-ray scatter, something that Siddon's method does not model.

Compared to previous neural field methods, traditional CBCT reconstruction algorithms have much faster runtimes. The slowest traditional method, Nesterov, takes no more than \SI{5}{\minute} for any number of views~(\cref{fig:quantitative}C). In contrast, the two slowest methods overall, SAX-NeRF and NAF, take on average \SI{10.3}{\hour} and \SI{30.6}{\hour}, respectively, to reconstruct a volume from 60 input views. While previous neural fields with hashgrid-augmented neural architectures claim to reconstruct CBCT volumes in tens of minutes~\cite{zha2022naf}, these runtimes are achieved when testing on low-resolution synthetic X-rays rendered from subsampled CT scans. Evaluations on real X-rays acquired at a standard resolution for clinical scanners demonstrate that existing neural fields are prohibitively slow for real-world deployment. Comparatively, reconstruction with Siddon's method and trilinear interpolation in DiffVox is an order of magnitude faster than existing neural fields, requiring only \SI{45.1}{\min} and \SI{39.1}{\min} for 60-view reconstruction, respectively. 

\begin{table}[t]
\centering
\caption{Quantitative comparison for 3D reconstruction from 15 views averaged over the 40 test walnuts. Our methods excel across multiple image fidelity metrics, outperforming traditional and neural CBCT reconstruction methods. Additionally, our voxel-based reconstruction method is an order of magnitude faster than existing neural field approaches. Metrics are reported as \textit{mean (se)}.}
\resizebox{\textwidth}{!}{
    \begin{tabular}{lccccc}
    \hline
                                        & SSIM ($\uparrow$)      & PSNR ($\uparrow$)     & MSE $\times 10^{-5}$($\downarrow$) & PCC ($\uparrow$)     & Runtime ($\downarrow$) \\ \hline
    FDK~\cite{feldkamp1984practical}    & 0.152 (0.014)          & 21.04 (0.67)          & 58.9 (0.62)          & 0.47 (0.01)          & \textbf{0.94 sec} \\
    CGLS~\cite{landman2023krylov}       & 0.505 (0.024)          & 30.01 (0.67)          & 7.47 (0.08)          & 0.76 (0.01)          & \uline{2.36 sec}          \\
    SIRT~\cite{gregor2008computational} & 0.809 (0.013)          & 32.99 (0.70)          & 3.87 (0.16)          & 0.88 (0.03)          & 34.6 sec         \\
    Nesterov~\cite{der2019cone}         & 0.834 (0.011)          & 34.35 (0.69)          & 2.81 (0.09)          & 0.92 (0.02)          & 2.3 min \\
    NAF~\cite{zha2022naf}               & 0.782 (0.015)          & 31.72 (0.69)          & 5.09 (0.13)          & 0.85 (0.02)          & 123.5 min        \\
    SAX-NeRF~\cite{cai2024structure}    & 0.835 (0.014)          & 33.43 (0.89)          & 3.97 (0.32)          & 0.88 (0.06)          & 160 min          \\
    \textbf{DiffVox -- Trilinear (Ours)}           & \uline{0.912 (0.006)}  & \uline{37.23 (0.67)}  & \uline{1.43 (0.03)}  & \textbf{0.96 (0.00)} & 10.0 min         \\
    \textbf{DiffVox -- Siddon's (Ours)}              & \textbf{0.923 (0.006)} & \textbf{37.69 (0.67)} & \textbf{1.28 (0.03)} & \textbf{0.96 (0.00)} & 11.3 min         \\ \hline
    \textbf{DiffVox -- Trilinear w/o TV (Ours)}    & 0.827 (0.008)          & 29.25 (0.56)          & 10.26 (0.15)           & 0.73 (0.01)          & 9.7 min          \\
    \textbf{DiffVox -- Siddon's w/o TV (Ours)}       & 0.812 (0.007)          & 27.41 (0.56)          & 15.72 (0.29)           & 0.67 (0.01)          & 10.1 min         \\ \hline
\end{tabular}
}
\label{tab:quant}
\vspace{-1em}
\end{table}

\section{Conclusions}
\label{sec:conclusions}
We present DiffVox, a modular framework for performing CBCT reconstruction with differentiable X-ray rendering. In a thorough analysis of a dataset with real X-rays, we found that DiffVox achieved high quality reconstructions given an extremely sparse number of input views ($\leq 20$), a task for which traditional and neural CBCT reconstruction algorithms proved insufficient. In particular, we observed that Siddon's method achieved slightly higher accuracy than trilinear interpolation at the expense of moderately increased runtimes. In addition to achieving higher fidelity 3D reconstructions and 2D novel views than existing algorithms, DiffVox was an order of magnitude faster to train than previous neural field methods: DiffVox required tens of minutes to reconstruct clinical-resolution CBCT scans whereas previous methods required hours to days. Furthermore, compared to continuous neural representations, parameterizing the unknown volume as a discrete voxelgrid endowed unique advantages such as fast convergence and the ability to utilize well-validated reconstruction regularizers. Finally, this analysis demonstrated the importance of evaluating reconstruction algorithms for clinical use on real images instead of primarily utilizing synthetic data. 

Our preliminary results naturally suggest several potential areas for future work. For example, incorporating a more realistic implementation of X-ray image formation (e.g., Siddon's method) into neural fields may yield compounding benefits. Additionally, as our physics-based reconstruction framework is fully differentiable, voxelgrid priors beyond TV regularization can be easily incorporated for specialized applications, such as vascular reconstruction~\cite{frangi1998multiscale}. To further exploit the differentiability of our X-ray renderer, it is also possible to jointly optimize the acquisition geometry of the X-ray images along with the unknown volume to address inaccurate or unspecified camera poses~\cite{ruckert2022neat, gopalakrishnan2024intraoperative}. Additionally, instead of initializing our voxelgrid with all zeros, one could instead initialize with the volume estimated by Nesterov or SIRT. Further refining this initial estimate with differentiable rendering could sufficiently reduce the runtime of DiffVox to the point of clinical relevance while retaining its superior reconstruction quality. Note that this initialization scheme is a unique advantage of our discrete voxelgrid representation, and not possible with continuous neural parameterizations.

\textbf{Acknowledgments.}
This work was supported in part by NIH NIBIB NAC P41EB015902, NIH NINDS U19NS115388, NIH NIBIB 5T32EB001680-19, and NIH R01EB034223.

\bibliography{references}

\begin{thebibliography}{17}
\providecommand{\natexlab}[1]{#1}
\providecommand{\url}[1]{\texttt{#1}}
\expandafter\ifx\csname urlstyle\endcsname\relax
  \providecommand{\doi}[1]{doi: #1}\else
  \providecommand{\doi}{doi: \begingroup \urlstyle{rm}\Url}\fi

\bibitem[Feldkamp et~al.(1984)Feldkamp, Davis, and Kress]{feldkamp1984practical}
Lee~A Feldkamp, Lloyd~C Davis, and James~W Kress.
\newblock Practical cone-beam algorithm.
\newblock \emph{Journal of the Optical Society of America A}, 1\penalty0 (6):\penalty0 612--619, 1984.

\bibitem[Landman et~al.(2023)Landman, Biguri, Hatamikia, Boardman, Aston, and Sch{\"o}nlieb]{landman2023krylov}
Malena~Sabat{\'e} Landman, Ander Biguri, Sepideh Hatamikia, Richard Boardman, John Aston, and Carola-Bibiane Sch{\"o}nlieb.
\newblock On {Krylov} methods for large-scale {CBCT} reconstruction.
\newblock \emph{Physics in Medicine \& Biology}, 68\penalty0 (15):\penalty0 155008, 2023.

\bibitem[Gregor and Benson(2008)]{gregor2008computational}
Jens Gregor and Thomas Benson.
\newblock Computational analysis and improvement of {SIRT}.
\newblock \emph{IEEE Transactions on Medical Imaging}, 27\penalty0 (7):\penalty0 918--924, 2008.

\bibitem[Zha et~al.(2022)Zha, Zhang, and Li]{zha2022naf}
Ruyi Zha, Yanhao Zhang, and Hongdong Li.
\newblock {NAF}: neural attenuation fields for sparse-view {CBCT} reconstruction.
\newblock In \emph{International Conference on Medical Image Computing and Computer-Assisted Intervention}, pages 442--452. Springer, 2022.

\bibitem[R{\"u}ckert et~al.(2022)R{\"u}ckert, Wang, Li, Idoughi, and Heidrich]{ruckert2022neat}
Darius R{\"u}ckert, Yuanhao Wang, Rui Li, Ramzi Idoughi, and Wolfgang Heidrich.
\newblock {NeAT}: Neural adaptive tomography.
\newblock \emph{ACM Transactions on Graphics (TOG)}, 41\penalty0 (4):\penalty0 1--13, 2022.

\bibitem[Lin et~al.(2023)Lin, Luo, Zhao, and Li]{lin2023learning}
Yiqun Lin, Zhongjin Luo, Wei Zhao, and Xiaomeng Li.
\newblock Learning deep intensity field for extremely sparse-view {CBCT} reconstruction.
\newblock In \emph{International Conference on Medical Image Computing and Computer-Assisted Intervention}, pages 13--23. Springer, 2023.

\bibitem[M{\"u}ller et~al.(2022)M{\"u}ller, Evans, Schied, and Keller]{muller2022instant}
Thomas M{\"u}ller, Alex Evans, Christoph Schied, and Alexander Keller.
\newblock Instant neural graphics primitives with a multiresolution hash encoding.
\newblock \emph{ACM Transactions on Graphics (TOG)}, 41\penalty0 (4):\penalty0 1--15, 2022.

\bibitem[Cai et~al.(2024)Cai, Wang, Yuille, Zhou, and Wang]{cai2024structure}
Yuanhao Cai, Jiahao Wang, Alan Yuille, Zongwei Zhou, and Angtian Wang.
\newblock Structure-aware sparse-view {X}-ray {3D} reconstruction.
\newblock In \emph{Proceedings of the IEEE/CVF Conference on Computer Vision and Pattern Recognition}, pages 11174--11183, 2024.

\bibitem[Sun et~al.(2022)Sun, Sun, and Chen]{sun2022direct}
Cheng Sun, Min Sun, and Hwann-Tzong Chen.
\newblock Direct voxel grid optimization: Super-fast convergence for radiance fields reconstruction.
\newblock In \emph{Proceedings of the IEEE/CVF Conference on Computer Vision and Pattern Recognition}, pages 5459--5469, 2022.

\bibitem[Gopalakrishnan and Golland(2022)]{gopalakrishnan2022fast}
Vivek Gopalakrishnan and Polina Golland.
\newblock Fast auto-differentiable digitally reconstructed radiographs for solving inverse problems in intraoperative imaging.
\newblock In \emph{Workshop on Clinical Image-Based Procedures}, pages 1--11. Springer, 2022.

\bibitem[Rudin et~al.(1992)Rudin, Osher, and Fatemi]{rudin1992nonlinear}
Leonid~I Rudin, Stanley Osher, and Emad Fatemi.
\newblock Nonlinear total variation based noise removal algorithms.
\newblock \emph{Physica D: Nonlinear Phenomena}, 60\penalty0 (1-4):\penalty0 259--268, 1992.

\bibitem[Kingma and Ba(2015)]{kingma2014adam}
Diederik~P. Kingma and Jimmy Ba.
\newblock Adam: {A} method for stochastic optimization.
\newblock In \emph{3rd International Conference on Learning Representations, {ICLR} 2015, San Diego, CA, USA, May 7-9, 2015, Conference Track Proceedings}, 2015.

\bibitem[Siddon(1985)]{siddon1985fast}
Robert~L Siddon.
\newblock Fast calculation of the exact radiological path for a three-dimensional {CT} array.
\newblock \emph{Medical Physics}, 12\penalty0 (2):\penalty0 252--255, 1985.

\bibitem[Der~Sarkissian et~al.(2019)Der~Sarkissian, Lucka, van Eijnatten, Colacicco, Coban, and Batenburg]{der2019cone}
Henri Der~Sarkissian, Felix Lucka, Maureen van Eijnatten, Giulia Colacicco, Sophia~Bethany Coban, and Kees~Joost Batenburg.
\newblock A cone-beam {X}-ray computed tomography data collection designed for machine learning.
\newblock \emph{Scientific Data}, 6\penalty0 (1):\penalty0 215, 2019.

\bibitem[Van~Aarle et~al.(2016)Van~Aarle, Palenstijn, Cant, Janssens, Bleichrodt, Dabravolski, De~Beenhouwer, Joost~Batenburg, and Sijbers]{van2016fast}
Wim Van~Aarle, Willem~Jan Palenstijn, Jeroen Cant, Eline Janssens, Folkert Bleichrodt, Andrei Dabravolski, Jan De~Beenhouwer, K~Joost~Batenburg, and Jan Sijbers.
\newblock Fast and flexible {X}-ray tomography using the {ASTRA} toolbox.
\newblock \emph{Optics Express}, 24\penalty0 (22):\penalty0 25129--25147, 2016.

\bibitem[Frangi et~al.(1998)Frangi, Niessen, Vincken, and Viergever]{frangi1998multiscale}
Alejandro~F Frangi, Wiro~J Niessen, Koen~L Vincken, and Max~A Viergever.
\newblock Multiscale vessel enhancement filtering.
\newblock In \emph{Medical Image Computing and Computer-Assisted Intervention—MICCAI’98: First International Conference Cambridge, MA, USA, October 11--13, 1998 Proceedings 1}, pages 130--137. Springer, 1998.

\bibitem[Gopalakrishnan et~al.(2024)Gopalakrishnan, Dey, and Golland]{gopalakrishnan2024intraoperative}
Vivek Gopalakrishnan, Neel Dey, and Polina Golland.
\newblock Intraoperative {2D/3D} image registration via differentiable {X}-ray rendering.
\newblock In \emph{Proceedings of the IEEE/CVF Conference on Computer Vision and Pattern Recognition}, pages 11662--11672, 2024.

\end{thebibliography}

\end{document}